\documentclass[preprint,11pt,aps,draft,nofootinbib,showpacs,showkeys]{revtex4}
\usepackage{amssymb}

\begin{document}

\title{Predictive statistical mechanics and macroscopic time evolution. A model for closed Hamiltonian systems}

\author{Domagoj Kui\'{c}}
\email{dkuic@pmfst.hr} 

\affiliation{University of Split, Faculty of Science, N. Tesle 12, 21000 Split, Croatia}

\date{September 21th, 2015}

\begin{abstract}
Predictive statistical mechanics is a form of inference from available data, without additional assumptions, for predicting reproducible phenomena. By applying it to systems with Hamiltonian dynamics, a problem of predicting the macroscopic time evolution of the system in the case of incomplete information about the microscopic dynamics was considered. In the model of a closed Hamiltonian system (i.e. system that can exchange energy but not particles with the environment) that with the Liouville equation uses the concepts of information theory, analysis was conducted of the loss of correlation between the initial phase space paths and final microstates, and the related loss of information about the state of the system. It is demonstrated that applying the principle of maximum information entropy by maximizing the conditional information entropy, subject to the constraint given by the Liouville equation averaged over the phase space, leads to a definition of the rate of change of entropy without any additional assumptions. In the subsequent paper \cite{kuic2} this basic model is generalized further and brought into direct connection with the results of nonequilibrium theory. 
\end{abstract}

\keywords{maximum entropy principle, information theory, statistical mechanics, nonequilibrium theory, Hamiltonian dynamics, entropy production}

\maketitle

\section{Introduction}\label{eqI}

Foundations of predictive statistical mechanics were formulated by E. T. Jaynes in his well know papers \cite{jaynes1,jaynes2}. There he gave a full development of the results of equilibrium statistical mechanics and the formalism of Gibbs \cite{gibbs} as a form of statistical inference based on Shannon's concept of a measure of information \cite{shannon}. Shannon's measure of information is also known as information entropy, and in the interpretation given by Jaynes it is the correct measure of the ``amount of uncertainty" represented by a  probability distribution \cite{jaynes3}. Maximization of information entropy subject to given constraints is a central concept in Jaynes' approach known as the {\it principle of maximum information entropy}. Application of this principle allows the construction of a probability distribution which includes in the distribution only information represented by given constraints, without any additional assumptions. Jaynes' approach is based on the Gibbs' formalism of statistical mechanics, which Jaynes considered to represent a general method of statistical inference in different problems where available information is not complete \cite{jaynes5}. This includes equilibrium statistical mechanics \cite{jaynes1} and the formulation of a theory of irreversibility \cite{jaynes2}, that Jaynes tried to accomplish in his later works \cite{jaynes3,jaynes4,jaynes5,jaynes6,jaynes7}.         

Predictions and calculations for different irreversible processes usually involve three distinct stages \cite{jaynes5}: 
\begin{itemize}
\item[(1)] Setting up an ``ensemble", i.e., choosing an initial density matrix, or in our case an $N$-particle distribution, which is to describe our initial knowledge about the system of interest;
\item[(2)] Solving the dynamical problem; i.e., applying the microscopic equations of motion to obtain the time evolution of the system; 
\item[(3)] Extracting the final physical predictions from the time developed ensemble. 
\end{itemize}
As fully recognized by Jaynes, the availability of the general solution of stage (1) simplifies the complicated stage (2). The problem includes also an equally important stage (0) consisting of some kind of measurement or observation defining both the system and the problem \cite{grandy1}. In direct mathematical attempts that lead to a theory of irreversibility, the Liouville theorem with the conservation of phase space volume inherent to Hamiltonian dynamics, is often represented as one of the main difficulties. Relation of the Liouville equation and irreversible macroscopic behavior is one of the central problems in statistical mechanics. For that reason this extremely complicated equation is reduced to an irreversible equation known as Boltzmann equation, rate equation or master equation. On the other hand, Jaynes considers the Liouville equation and the related constancy in time of Gibbs' entropy as precisely the dynamical property needed for solution of this problem, considering it to be more of conceptual than mathematical nature \cite{jaynes3,jaynes4}. In the simple demonstration based on the Liouville theorem, this makes possible for Jaynes to generalize the second law beyond the restrictions of initial and final equilibrium states, by considering it a special case of a general restriction on the direction of any reproducible process \cite{jaynes4,jaynes8}. The real reason behind the second law, since phase space volume is conserved in the dynamical evolution, is a fundamental requirement on any reproducible process that the phase space volume $W^{\prime}$, compatible with the final (macroscopic) state, can not be less than the phase space volume $W_0$ which describes our ability to reproduce the initial state \cite{jaynes4}.

Mathematical clarity of Jaynes' viewpoint has its basis in a limit theorem noted by Shannon \cite{shannon}, known as the asymptotic equipartition theorem of information theory. Application of this theorem relates in certain cases, in a limit of large number of particles, the Boltzmann's formula for entropy of a macrostate and the Gibbs expression for entropy \cite{jaynes4,jaynes5,jaynes7}. Mathematical connection with the Boltzmann's interpretation of entropy as the logarithm of the number or ways (or microstates) by which a macroscopic state can be realized, gives then a simple physical interpretation to the Gibbs' formalism, and its generalization in the maximum-entropy formalism.  Maximization of the information entropy subject to given constraints then predicts the macroscopic behavior that can happen in the greatest number of ways compatible with the information represented by given constraints. In application to time dependent processes, this is referred to by Jaynes as the maximum caliber principle \cite{jaynes6,jaynes7}. Jaynes clearly stated that predictive statistical mechanics does not represent a physical theory that explains the behavior of different systems by deductive reasoning from the first principles, but a form of statistical inference that makes predictions of observable phenomena from incomplete information \cite{jaynes6}. For this reason predictive statistical mechanics can not claim certainty for its predictions in the way that a deductive theory can. This does not mean that predictive statistical mechanics ignores the laws of microphysics; it certainly uses everything known about the structure of microstates and any data on macroscopic quantities, without making any extra physical assumptions beyond what is given by available information. It is important to note that sharp, definite predictions of macroscopic behavior are possible only when certain behavior is characteristic of each of the overwhelming majority of microstates compatible with data. For the same reason, this is just the behavior that is reproduced experimentally under those constraints; this is known essentially as the principle of macroscopic uniformity \cite{jaynes1,jaynes2}, or the {\it principle of macroscopic reproducibility} \cite{jaynes7}. In somewhat different context this property is recognized as the concept of macroscopic determinism, whose precise definition involves some sort of thermodynamic limit \cite{harris,garrod}. The second law of thermodynamics predicts only that a change of macroscopic state will go in the general direction of greater final entropy, but not at which rate, or along which path \cite{jaynes6,jaynes7,jaynes8}. It is clear that better predictions are possible only by introducing more information. Macrostates of higher entropy can be realized in overwhelmingly more ways, and this is the basic reason for high reliability of the Gibbs equilibrium predictions \cite{jaynes7}. In this context, Jaynes also speculated that accidental success in the reversal of an irreversible process is exponentially improbable \cite{jaynes8}.   

Jaynes' interpretation of irreversibility and the second law reflects the point of view of the actual experimenter. Zurek \cite{zurek1} has proposed the definition of physical entropy as the sum of the missing information about the microscopic state, given by Shannon's information entropy, and the algorithmic information content present in the available data about the system. In the limit of Zurek's approach in which measurement is complete and the microstate is known, physical entropy of the system is given by the algorithmic information content about the microscopic state in which the system is found \cite{zurek1}. Zurek's interpretation of the physical entropy and thermodynamics is given at the level of observers that can acquire information through measurements and process it in accordance with the basic laws of computation in a manner analogous to Turing machines. Jaynes has maintained the position that measurements \cite{jaynes2} in practice always represent far less than the maximum observation which would enable us to determine a definite pure state (i.e. the microscopic state of the system). This is the reason why \cite{jaynes2} we must have recourse to maximum-entropy inference in order to represent our degree of knowledge about the system in a way free of arbitrary assumptions with regard to missing information.

MaxEnt algorithm is a general method of constructing the probability distribution by applying the principle of maximum information entropy in cases when distribution is not determined uniquely by available information. Arbitrary assumptions can be avoided by selecting the probability distribution which is compatible with the available information, and which is characterized by largest uncertainty related to missing information. Inferences drawn from such probability distribution depend only on a real degree of knowledge \cite{jaynes1,jaynes2}. Probability distribution that maximizes the information entropy (uncertainty) subject to constraints given by available macroscopic data, in predictive statistical mechanics represents real uncertainty related to missing information about the actual microscopic state of the system.     

In a similar line of reasoning Grandy \cite{grandy2,grandy3,grandy} has developed a detailed model of time dependent probabilities and density matrix for macroscopic systems with time dependent constraints within the MaxEnt formalism, and applied it to typical processes in nonequilibrium thermodynamics and hydrodynamics \cite{grandy4,grandy}. In a context of the interplay between macroscopic constraints on the system and its microscopic dynamics, it is interesting to note that MaxEnt formalism has been also studied as a method of approximately solving partial differential equations governing the time evolution of probability distributions. For more complete further reference, we only mention here that this method, among other examples, has been applied to the Liouville--von Neumann equation \cite{tishby}, the family of dynamical systems with divergenceless phase space flows including Hamiltonian systems \cite{plastino1}, the generalized Liouville equation and continuity equations \cite{plastino2}. Universality of this approach has been established for the general class of evolution equations that conform to the essential requirements of linearity and preservation of normalization \cite{plastino3}. This method has been also considered for classical evolution equations with source terms within a framework where normalization is not preserved \cite{plastino4}.  

In this and in the subsequent paper \cite{kuic2} we consider the application of predictive statistical mechanics on the problem of predicting the macroscopic time evolution of systems with Hamiltonian dynamics, in the case when the information about the microscopic dynamics of the system is not complete. For this purpose we have developed a basic theoretical model for a closed system with Hamiltonian dynamics. Concepts of Hamiltonian mechanics and probability distributions in the phase space applied in this model are defined in Sections \ref{secHD} and  \ref{sec_Mic}. In Section \ref{secIE} information entropies that correspond to those probability distributions are defined. The model is set and its results are analyzed in Section \ref{secAMFCHS}. Results that have already been presented in \cite{kuic} were obtained in a model of a closed system with the time independent Hamiltonian function. In this paper, we have included in this basic model also closed systems with Hamiltonian function that depends on time. Conclusions based on these results are presented in Section \ref{secCGARWNT}. They are the basis for further generalization of this basic theoretical model in the subsequent paper \cite{kuic2}, where it is brought in direct connection with the results of the nonequilibrium theory.

\section{Hamiltonian dynamics and phase space paths} \label{secHD} 
The dynamical state of a Hamiltonian system with $s$ degrees of freedom is described by the generalized coordinates $q_1, q_2, \dots , q_s$ and their conjugate momenta $p_1, p_2, \dots , p_s$. At any time $t$ it is represented by a point in the $2s$-dimensional space $\Gamma $ called the phase space of the system. The notation $(q,p) \equiv  (q_1, q_2, \dots , q_s, p_1, p_2, \dots , p_s)$ is introduced for the set of generalized coordinates and conjugate momenta forming together $2s$ coordinates of the phase space $\Gamma $. The time dependence of $2s$ dynamical variables $(q,p)$ is determined by Hamilton's equations  
\begin{equation}
\dot q_i = \frac{\partial H}{\partial p_i} , \qquad \qquad \dot p_i =  - \frac{\partial H}{\partial q_i} , \qquad \qquad i=1,2, \dots, s , \label{eq1}
\end{equation} 
where $H \equiv H(q,p,t)$ is the Hamiltonian function of the system. For the given values $(q_0,p_0)$ at some time $t_0$, Hamilton's equations (\ref{eq1}) and its solution uniquely determine the values of dynamical variables $(q,p)$ at any other time $t$: 
\begin{equation}
q_i = q_i(t; q_0,p_0) , \qquad \qquad p_i = p_i(t; q_0,p_0) , \qquad \qquad i=1,2, \dots, s . \label{eq2} 
\end{equation}   
Hence, a point $(q,p)$ in the phase space $\Gamma $ representing the state of the system describes over time a curve called a {\it phase space path}, uniquely determined by the solution of (\ref{eq1}). The set $\Omega (\Gamma )$ is the set of all phase space paths in $\Gamma $. At time $t$ through the point $(q,p)_\omega  \in \Gamma $ passes only one path $\omega \in \Omega (\Gamma )$, and this is denoted by the index in $(q,p)_\omega $, where $\omega \in \Omega (\Gamma )$.  The velocity $v$ of the point $(q,p)$ in the phase space $\Gamma $ corresponding to values of dynamical variables at time $t$ is given by 
\begin{equation}
v \equiv  \vert {\bf v}\vert  = \sqrt {\sum _{i=1}^{s} \left (\frac {dq_{i}}{dt}\right ) ^2 + \sum _{i=1}^{s}\left (\frac{dp_{i}}{dt}\right ) ^2} = \sqrt {\sum _{i=1}^{s} \left (\frac {\partial H}{\partial p_i}\right ) ^2 + \sum _{i=1}^{s}\left (\frac{\partial H}{\partial q_i}\right ) ^2}. \label{eq2a}
\end{equation}
The velocity vector ${\bf v}((q,p)_\omega , t)$ is tangential at the point  $(q,p)_\omega  \in \Gamma $ to the phase space path $\omega $ passing through it at time $t$. This defines the velocity vector field ${\bf v}((q,p), t)$ on $\Gamma $.

\section{Microstate probability and path probability} \label{sec_Mic}
It is now possible to relate the microstate probability and the path probability in the phase space $\Gamma $ of the system. Let the function $f(q,p,t)$ be a microstate probability density function on $\Gamma $. All points in the phase space $\Gamma $ move according to Hamilton's equations (\ref{eq1}) and $f(q,p,t)$ satisfies the Liouville equation
\begin{equation}
{\partial f \over \partial t} + \sum _{i=1} ^s \left ({\partial f \over \partial q_i}{\partial H \over \partial p_i} - {\partial f \over \partial p_i}{\partial H \over \partial q_i}\right ) \equiv {df \over dt} = 0 .\label{eq9}
\end{equation}   
Since $df/dt$ is a total or hydrodynamic derivative, (\ref{eq9}) expresses that the time rate of change of $f(q,p,t)$ is zero along any phase space path given by the solution of Hamilton's equations. In the notation used here, this fact is written as
\begin{equation}
f((q,p)_\omega ,t) = f((q_0,p_0)_\omega ,t_0) , \label{eq10}
\end{equation}
where points on the path $\omega \in \Omega (\Gamma )$ are related by (\ref{eq2}).

In order to relate the microstate probability and the path probability in the phase space $\Gamma $, probability density function $\mathcal {F}(q,p,t;q_0,p_0 ,t_0)$ is introduced on the $4s$-dimensional space $\Gamma \times \Gamma $. This function has the following special properties. If the integral of $\mathcal {F}(q,p,t;q_0,p_0 ,t_0)$ is taken over the phase space $\Gamma $ with $(q,p)$ as the integration variables, it gives the microstate probability density function $f(q_0,p_0,t_0)$ at time $t_0$, 
\begin{equation}
f(q_0,p_0, t_0) = \int _\Gamma \mathcal{F}(q,p,t;q_0,p_0 ,t_0) d\Gamma .  \label{eq11}
\end{equation}
Microstate probability density function $f(q,p,t)$ at time $t$ is obtained analogously,
\begin{equation}
f(q,p, t) = \int _\Gamma \mathcal{F}(q,p,t;q_0,p_0 ,t_0) d\Gamma_0 . \label{eq12}
\end{equation}
It is straightforward to prove, using relation (\ref{eq10}), that (\ref{eq11}) and (\ref{eq12}) are satisfied if the function $\mathcal{F}(q,p,t;q_0,p_0 ,t_0)$ has the following form:
\begin{equation}
\mathcal{F}(q,p,t; q_0,p_0 ,t_0) = f(q,p, t)\prod_{i=1}^{s}\delta (q_i - q_i(t;q_0,p_0))\delta (p_i - p_i(t;q_0,p_0)) , \label{eq13}
\end{equation}
where $q_i(t;q_0,p_0)$ and $p_i(t;q_0,p_0)$ are given by (\ref{eq2}) and $\delta$-s are Dirac delta functions. In the space $\Gamma \times \Gamma $ function $\mathcal{F}(q,p,t;q_0,p_0 ,t_0)$ given by (\ref{eq13}) represents the probability density that the point corresponding to the state of the system is in the element $d\Gamma _0$ around the point $(q_0,p_0)$ at time $t_0$ and in the element $d\Gamma $ around the point $(q,p)$ at time $t$.

\subsection{Time independent Hamiltonian function}

Now, we assume that the set $M$ of all points in $\Gamma $ that represent possible microstates of the system is invariant to Hamiltonian motion. We also assume that the Hamiltonian function does not depend on time $H = H(q,p)$ . The invariance of the measure $d\Gamma $ to Hamiltonian motion and the fact that the velocity field ${\bf v}(q,p)$ in $\Gamma $ is stationary as the Hamiltonian function does not depend on time, lead to the following consequence. For any phase space path $\omega \in \Omega (\Gamma )$, the product of the velocity $v((q,p)_\omega )$ and the infinitesimal element $dS_{\omega }$ of the surface intersecting the path $\omega $ perpendicularly at the point $(q,p)_\omega $, is constant under Hamiltonian motion along the entire length of the path $\omega $, i.e., 
\begin{equation}
v ((q,p)_\omega ) dS_\omega  = const . \label{eq4b}
\end{equation}
For any two points $(q_0,p_0)_\omega $ and $(q_a,p_a)_\omega $ on the same path $\omega $, the following relation is obtained from (\ref{eq4b}):
\begin{equation}
v ((q_0,p_0)_\omega ) dS_{0 \omega}  = v ((q_a,p_a)_\omega ) dS_{a \omega} . \label{eq4c}
\end{equation}
The infinitesimal element $dS_{0 \omega }$ of the surface $S_0(M)$ intersects the path $\omega $ perpendicularly at the point $(q_0,p_0)_\omega$. The surface $S_0(M)$ is perpendicular to all paths in the set $\Omega (M)$ of paths in $M$. The infinitesimal element $dS_{a \omega }$ of the surface $S_a(M)$ intersects the path $\omega $ perpendicularly at the point $(q_a,p_a)_\omega$. Like the surface $S_0(M)$, surface $S_a(M)$ is also perpendicular to all paths in $\Omega (M)$. The infinitesimal elements $dS_{0 \omega }$ and $dS_{a \omega}$ of the two surfaces $S_0(M)$ and $S_a(M)$ are connected by the path $\omega $ and neighboring paths determined by solutions of Hamilton's equations. The integral over surface $S_a(M)$ is transformed using (\ref{eq4c}) into integration over surface $S_0(M)$,
\begin{equation}
\int _{S_a(M)} dS _{a \omega } = \int _{S_0(M)} \frac{v((q_0,p_0)_\omega )}{v ((q_a,p_a)_\omega)} dS _{0 \omega } .  \label{eq4d}
\end{equation}
Functional dependence between the points $(q_0,p_0)_\omega $ and $(q_a,p_a)_\omega $ on the path $\omega $ is  not explicitly written in the integral (\ref{eq4d}); it is implied that this functional dependence is determined from solutions of Hamilton's equations and the additional condition of perpendicularity of the surfaces $S_0(M)$ and $S_a(M)$ to all paths in $\Omega (M)$. It is important to emphasize that perpendicularity of the surfaces $S_0(M)$ and $S_a(M)$ to all phase space paths in $\Omega (M)$ is implied by the definition of these surfaces and not as a consequence of Hamiltonian time evolution. It is also clear that the measure defined on the surface $S_0(M) $ can be utilized as a measure on the set $\Omega (M)$ of all phase space paths in some invariant set $M $. The correspondence between points  $(q_0,p_0)_\omega  \in S_0(M) $ and paths $\omega \in \Omega (M)$ is one-to-one.

The infinitesimal volume element $d\Gamma _a$ around the point $(q_a,p_a)_\omega$ through which the path $\omega \in \Omega (M)$ passes can be written as $d\Gamma _a = ds_{a \omega } dS_{a \omega }$. Here, $ds_{a \omega}$ is the infinitesimal distance along the path $\omega $, i.e. the infinitesimal arc length element of the path $\omega $. The integral (\ref{eq12}) can now be written as  
\begin{eqnarray}
f(q,p, t) & = & \int _\Gamma \mathcal{F}(q,p,t;q_a,p_a ,t_0) ds_{a \omega } dS_{a \omega } \nonumber\\
 & = & \int _{S_0(M)}dS_{0 \omega }v((q_0,p_0)_\omega )\int _\omega  \frac{\mathcal{F}(q,p,t;q_a,p_a ,t_0)}{v (q_a,p_a)} ds_{a \omega } , \label{eq15}
\end{eqnarray}
where in the first line $d\Gamma _a = ds_{a \omega } dS_{a \omega }$ (with dummy indices) is introduced and in the second line the integral is transformed in accordance with (\ref{eq4d}). Along with the lines leading to (\ref{eq15}), the function $G(q,p,t; (q_0,p_0)_\omega ,t_0 )$ is also introduced:
\begin{equation}
G(q,p,t; (q_0,p_0)_\omega ,t_0 ) = v((q_0,p_0)_\omega )\int _\omega  \frac{\mathcal{F}(q,p,t; q_a,p_a ,t_0)}{v (q_a,p_a)} ds_{a \omega } . \label{eq16}
\end{equation}
The integral in the definition of $G(q,p,t; (q_0,p_0)_\omega ,t_0 )$ in (\ref{eq16}) is over the entire length of the phase space path $\omega $ intersected perpendicularly by the surface $S_0(M)$ at the point $(q_0,p_0)_\omega$. Using (\ref{eq16}), relation (\ref{eq15}) is then written as  
\begin{equation}
f(q,p, t) = \int _{S_0(M)}G(q,p,t; (q_0,p_0)_\omega ,t_0 )dS_0 .  \label{eq17}
\end{equation}
It is clear that the expression 
\begin{equation}
G(q,p,t; (q_0,p_0)_\omega ,t_0 )dS_0d\Gamma \equiv d\mathrm{P}(q,p , t  \cap  (q_0,p_0)_\omega , t_0) , \label{eq18}
\end{equation}
represents the probability that the point corresponding to the state of the system at time $t_0$ is anywhere along the paths which pass through an infinitesimal element $dS_0$ around $(q_0,p_0)$ on the surface $S_0(M)$, and that at some different time $t$ it is in the volume element $d\Gamma $ around $(q,p)$.  Therefore, $G(q,p,t;(q_0,p_0)_\omega ,t_0)$ is a {\it joint probability density} of two continuous multidimensional variables, $(q,p)$ in $\Gamma $ and $(q_0,p_0)_\omega$ in $S_0(M)$. 

With (\ref{eq18}) and the definition of the joint density $G(q,p,t;(q_0,p_0)_\omega ,t_0)$, the definition of the path probability density $F((q_0,p_0)_\omega ,t_0)$ is now straightforward. It is given by the integral  
\begin{equation}
F((q_0,p_0)_\omega ,t_0) = \int_\Gamma  G(q,p,t; (q_0,p_0)_\omega ,t_0 )d\Gamma . \label{eq19}
\end{equation}
Then, in accordance with the theory of probability, the ratio 
\begin{equation}
{G(q,p,t;(q_0,p_0)_\omega ,t_0) dS_0 d\Gamma \over F((q_0,p_0)_\omega ,t_0) dS _0} \equiv d\mathrm{P} (q,p , t \vert (q_0,p_0)_\omega , t_0), \label{eq21} 
\end{equation}
represents the {\it conditional probability} that at time $t$ the point corresponding to the state of the system is in the element $d\Gamma $ around $(q,p)$, if at time $t_0$ it is anywhere along the paths passing through the infinitesimal element $dS_0$ around $(q_0,p_0)$ on the surface $S_0(M)$. Relation (\ref{eq19}) then proves that the integral of (\ref{eq21}) over $\Gamma $ satisfies the normalization condition, i.e.,
\begin{equation}
\int _\Gamma { G(q,p,t;(q_0,p_0)_\omega ,t_0) \over F((q_0,p_0)_\omega ,t_0) }d\Gamma   = 1 . \label{eq22} 
\end{equation}
The conditional probability density $D(q,p , t \vert (q_0,p_0)_\omega , t_0)$ that corresponds to conditional probability (\ref{eq21}) is defined by the relation
\begin{equation}
D(q,p , t \vert (q_0,p_0)_\omega , t_0) = {G(q,p,t;(q_0,p_0)_\omega ,t_0) \over F((q_0,p_0)_\omega ,t_0) } . \label{eq23} 
\end{equation}

The relation (\ref{eq21}), like the relation (\ref{eq18}), represents probability which is conserved in the phase space $\Gamma $. The total time derivative of the probability (\ref{eq21}), i.e. its time rate of change along the Hamiltonian flow lines, is equal to zero. In the relation (\ref{eq21}), the path probability density $F((q_0,p_0)_\omega ,t_0)$ and the surface element $dS _0$ are independent of the variables $t$ and $(q,p)$. Also, measure $d\Gamma $ is invariant to Hamiltonian motion. Therefore, the total time derivative of the conditional probability (\ref{eq21}) is equal to zero if and only if 
\begin{equation}
{dG \over dt} \equiv {\partial G \over \partial t} + \sum _{i=1} ^s \left ({\partial G \over \partial q_i}{\partial H \over \partial p_i} - {\partial G \over \partial p_i}{\partial H \over \partial q_i}\right )  = 0\ . \label{eq24}
\end{equation} 
This is a straightforward demonstration that the joint density $G(q,p,t;(q_0,p_0)_\omega ,t_0)$ satisfies the equation analogous to the Liouville equation (\ref{eq9}) for the microstate probability density $f(q,p,t)$.

\subsection{Time dependent Hamiltonian function}
If the Hamiltonian function $H = H(q,p,t)$ and Hamilton's equations (\ref{eq1}) depend on time, the subsequent and precedent motion in the phase space $\Gamma $ depends on the choice of the initial moment of time $t_0$. For the initial values $(q_0,p_0)$ given at time $t_0$, Hamilton's equations and its solution (\ref{eq2}) uniquely determine the phase space path which passes through $(q_0,p_0)$ at time $t_0$, and thus determine the points corresponding to subsequent and precedent values of the dynamical variables $(q,p)$. For the same initial values $(q_0,p_0)$ given at time $t^\prime_0 \neq t_0$, Hamilton's equations and its solution uniquely determine the phase space path which passes through $(q_0,p_0)$ at time $t^\prime_0$. If Hamilton's equations depend on time then these two phase space paths may be different. Invariance of Hamilton's equations to time translations is disrupted and, as a result, phase space paths are no longer time independent objects. This means that through the same point in $\Gamma $ at two different moments of time two different phase space paths may pass.  

This is an important distinction compared to the case of time independent Hamiltonian described in the previous subsection. If Hamiltonian function $H = H(q,p,t)$ and Hamilton's equations depend on time, unique specification of the phase space path requires the specification of the point through which the path passes and also the moment of time at which it is passing through that point. Because of this, in this case we can really say that the microstate probability density $f(q,p,t)$ represents the probability density of paths in the phase space $\Gamma $ at time $t$. Furthermore, by comparing (\ref{eq11}) and (\ref{eq19}), we see that the joint density $\mathcal{F}(q,p,t; q_0,p_0 ,t_0)$ now has the same interpretation that has been given to the joint density $G(q,p,t;(q_0,p_0)_\omega ,t_0)$ in the case of time independent Hamiltonian function. Accordingly, and in analogy with (\ref{eq21}), the expression
\begin{equation}
{\mathcal{F}(q,p,t; q_0,p_0 ,t_0) d\Gamma _0 d\Gamma \over f(q_0,p_0,t_0) d\Gamma _0} \equiv d\mathrm{P} (q,p , t \vert q_0,p_0 , t_0) , \label{eq24a}
\end{equation}
is the conditional probability that at time $t$ the point corresponding to the state of the system is in the element $d\Gamma $ around $(q,p)$, if at time $t_0$ it is in the element $d\Gamma _0$ around $(q_0,p_0)$ and therefore on the paths passing through it. The conditional probability density $B(q,p , t \vert q_0,p_0, t_0)$ that corresponds to the conditional probability (\ref{eq24a}) is defined by the relation
\begin{equation}
B(q,p , t \vert q_0,p_0, t_0) = {\mathcal{F}(q,p,t; q_0,p_0 ,t_0) \over f(q_0,p_0,t_0) } . \label{eq24b}
\end{equation} 
Using (\ref{eq10}), (\ref{eq13}) and (\ref{eq24b}), it is easy to see that
\begin{equation}
B(q,p , t \vert q_0,p_0, t_0) = \prod_{i=1}^{s}\delta (q_i - q_i(t;q_0,p_0))\delta (p_i - p_i(t;q_0,p_0)) . 
\end{equation} 
By demonstration analogous to that which lead to (\ref{eq24}), now applying it to the conditional probability (\ref{eq24a}), it is simple to show that the joint density $\mathcal{F}(q,p,t; q_0,p_0 ,t_0)$ also satisfies the Liouville equation, i.e. that
\begin{equation}
{d\mathcal{F} \over dt} \equiv {\partial \mathcal{F} \over \partial t} + \sum _{i=1} ^s \left ({\partial \mathcal{F} \over \partial q_i}{\partial H \over \partial p_i} - {\partial \mathcal{F} \over \partial p_i}{\partial H \over \partial q_i}\right )  = 0 . \label{eq24c}
\end{equation} 
To conclude, if Hamilton's equations depend on time then phase space paths are no longer properly specified only by the points through which they pass, time is also a necessary part of their specification. In order to take that into account in a sensible way, we use in that case the joint density $\mathcal{F}(q,p,t; q_0,p_0 ,t_0)$ given by (\ref{eq13}), and not the joint density $G(q,p,t;(q_0,p_0)_\omega ,t_0)$ whose definition (\ref{eq16}) was given for the case of time independent Hamiltonian function.

\section{Information entropies} \label{secIE}

In Shannon's information theory \cite{shannon} the quantity of the form  
\begin{equation}
H(p_1, \dots, p_n) = -K\sum_{i=1}^n p_i\log p_i , \label{eq24d}
\end{equation}
has a central role of measure of information, choice and uncertainty for different probability distributions $p_1, \dots, p_n$. From the understanding that the problem of constructing a communication device depends on the statistical structure of the information that is to be communicated (it depends for example on the probabilities $p_1, p_2, \dots, p_n$ of the symbols $A_1, A_2, \dots, A_n$ of some alphabet) Shannon gave until that time most general definition (\ref{eq24d}) of the measure of amount of information. Sequences of symbols or "letters" may form the set of "words" of certain length, and the amount of information is measured analogously. Positive constant $K$ in (\ref{eq24d}) depends on the choice of a unit for amount of information. In real applications expression (\ref{eq24d}), with logarithmic base $2$ and $K=1$, represents the expected number of bits per symbol necessary to encode the random signal forming a memoryless source. But perhaps it is most important that Shannon's interpretation of the function (\ref{eq24d}) is not dependent on the specific context of information theory. He defined the function (\ref{eq24d}) as a measure of {\itshape our uncertainty} related to the occurrence of possible events, or more specifically, as a measure of uncertainty {\itshape represented by the probability distribution} $p_1, p_2, \dots, p_n$. This is substantiated by three reasonable properties that are required from such a measure $H(p_1, \dots, p_n)$: continuity, monotonic increase with number of possibilities in case when all probabilities are equal, and the unique and consistent composition law for the addition of uncertainties when mutually exclusive events are grouped into composite events. These three properties,  as demonstrated by Shannon in his famous theorem, are sufficient to uniquely determine the form of the function $H(p_1, \dots, p_n)$ and it is given by (\ref{eq24d}). Shannon called the function (\ref{eq24d}) the entropy of the set of probabilities $p_1, p_2, \dots, p_n$.

In an analogous manner Shannon  has defined entropy of a continuous distribution and entropy of $N$-dimensional continuous distribution. Jaynes \cite{jaynes3}, on the other hand, deduced that the quantity
\begin{equation}
S_I = -\int w(x)\log\left [\frac{w(x)}{m(x)}\right ]dx . \label{eq25}
\end{equation}
corresponds to the quantity $-\sum_{i=1}^{n} p_i\log p_i$ for a discrete probability distribution $p_i$ which in a limit of infinite number of points tends to continuous distribution with the density function $w(x)$ (in such a way that the density of points, divided by their total number, approaches a definite function $m(x)$). Under a change of variables $w(x)$ and $m(x)$ transform in the same way, and the described limit process from a discrete to a continuous distribution, with the definition of measure function $m(x)$, yields the invariant information measure (\ref{eq25}). Invariance of the entropy of a continuous distribution under a change of the independent variable is thus achieved with a modification that follows from the mathematical deduction conducted by Jaynes, and this is readily generalized to the case when a discrete distribution passes to a continuous multidimensional distribution \cite{jaynes3}. If uniform measure $m=const.$ is assumed, then (\ref{eq25}) differs from Shannon's definition of entropy of a continuous distribution by an irrelevant additive constant. For example, in the quasiclassical limit of quantum statistical mechanics justification for this assumption is given by the standard proposition that each discrete quantum state corresponds to a volume $h^{3N}$ of the classical phase space.

Shannon \cite{shannon} has also defined joint and conditional entropies of a joint distribution of two continuous variables (which may themselves be multidimensional). In the previous section, joint probability density $G(q,p,t;(q_0,p_0)_\omega ,t_0)$ of two continuous multidimensional variables $(q,p)$ in $\Gamma $ and $(q_0,p_0)_\omega$ in $S_0(M)$ was introduced. Following the detailed explanation of (\ref{eq18}), $G(q,p,t;(q_0,p_0)_\omega ,t_0)dS_0d\Gamma $ represents the probability of the joint occurrence of two events: the first occurring at time $t_0$ on the set $\Omega (M)$ of all possible phase space paths and the second occurring at time $t$ on the set $M$ of all possible phase space points, the set which is invariant to Hamiltonian motion. As discussed in the previous section, in the case when Hamilton's equations depend on time, the same interpretation and role is given to the joint probability density $\mathcal{F}(q,p,t; q_0,p_0 ,t_0)$ of two continuous multidimensional variables in $\Gamma \times \Gamma $;  $(q_0,p_0)$ which corresponds to time $t_0$ and $(q,p)$ corresponding to time $t$. 

In accordance with Shannon's definition \cite{shannon}, {\it joint information entropy} of $G(q,p,t;(q_0,p_0)_\omega ,t_0)$ is given by
\begin{equation}
S_{I}^{G}(t, t_0) = - \int _{S_0(M)} \int_\Gamma G \log G \ d\Gamma dS _0  . \label{eq26}
\end{equation} 
The notation $S_{I}^{G}(t, t_0)$ indicates that it is a function of time $t$ and $t_0$, through the joint probability density $G \equiv  G(q,p,t;(q_0,p_0)_\omega ,t_0)$. Following Shannon's definition \cite{shannon}, {\it conditional information entropy} of $G(q,p,t;(q_0,p_0)_\omega ,t_0)$ is then given by
\begin{equation}
S_{I}^{DF} (t, t_0) = - \int _{S_0(M)} \int_\Gamma G\log \left[\frac{G}{F}\right ] \ d\Gamma dS _0  , \label{eq27}
\end{equation}
where $F \equiv  F((q_0,p_0)_\omega ,t_0 )$ is the path probability density (\ref{eq19}). Using the definition of $D(q,p,t | (q_0,p_0)_\omega ,t_0 )$ in (\ref{eq23}), one immediately obtains the equivalent form of the conditional information entropy (\ref{eq27}): 
\begin{equation}
S_{I}^{DF} (t, t_0) = - \int _{S_0(M)} \int_\Gamma DF\log D \ d\Gamma dS _0 . \label{eq28}
\end{equation}
From (\ref{eq28}) it is clear that the conditional information entropy $S_{I}^{DF} (t, t_0)$ is the average of the entropy of $D(q,p,t | (q_0,p_0)_\omega ,t_0 )$, weighted over all possible phase space paths $\omega \in \Omega (M)$ according to the path probability density $F((q_0,p_0)_\omega ,t_0 )$. Definitions of the joint $S_{I}^{\mathcal{F}}(t, t_0)$ and conditional $S_{I}^{Bf}(t, t_0)$ information entropies of the joint distribution with the density function $\mathcal{F}(q,p,t; q_0,p_0 ,t_0)$ are analogous to the definitions (\ref{eq26}), (\ref{eq27}) and (\ref{eq28}). There is no need to also write them here explicitly; they are readily obtained from (\ref{eq26}), (\ref{eq27}) and (\ref{eq28}), by changing the symbols with corresponding meanings as explained in the previous section: replace $G(q,p,t;(q_0,p_0)_\omega ,t_0)$ with $\mathcal{F}(q,p,t; q_0,p_0 ,t_0)$, $F((q_0,p_0)_\omega ,t_0 )$ with $f(q_0,p_0, t_0)$, $D(q,p , t \vert (q_0,p_0)_\omega , t_0)$ with $B(q,p , t \vert q_0,p_0 , t_0)$, $M $ and $S_0(M)$ with $\Gamma $, and $dS_0$ with $d\Gamma _0$.

Relation between the information entropies $S_{I}^{G}(t, t_0)$ and $S_{I}^{DF} (t, t_0)$, introduced in (\ref{eq26}) and (\ref{eq27}), is completed by introducing the information entropy of $F((q_0,p_0)_\omega ,t_0 )$, or alternatively, {\it path information entropy}: 
\begin{equation}
S_{I}^{F} (t_0) = - \int _{S_0(M)} F\log F \ dS _0 . \label{eq29} 
\end{equation}
Relation between $S_{I}^{G}(t, t_0)$, $S_{I}^{DF} (t, t_0)$ and $S_{I}^{F} (t_0)$ is obtained straightforwardly, using (\ref{eq23}) in (\ref{eq26}), and then applying the properties of probability distributions and definition (\ref{eq28}). In this way one obtains
\begin{equation}
S_{I}^{G}(t,t_0) = S_{I}^{DF}(t,t_0) + S_{I}^{F} (t_0) . \label{eq30}
\end{equation}
In accordance with \cite{shannon}, relation (\ref{eq30}) asserts that the uncertainty (or entropy) of the joint event is equal to the uncertainty of the first plus the uncertainty of the second event when the first is known. To be mathematically precise, uncertainty of a joint event means here the uncertainty of two random variables which are defined on the space of elementary events of the same probability space. Uncertainty of individual events is the uncertainty of these individual random variables. 

In general, uncertainty of the joint event is less then or equal to the sum of individual uncertainties, with the equality if (and only if) the two random variables are independent \cite{shannon}. The probability distribution of the joint event is given here by the density $G(q,p,t;(q_0,p_0)_\omega ,t_0)$. Information entropy or uncertainty of one of them (in this case called the second event because of its occurrence at a later time) is given by
\begin{equation}
S_{I}^{f} (t) = - \int_\Gamma f\log f \ d\Gamma . \label{eq31}
\end{equation}
The quantity $S_{I}^{f} (t)$ is the information entropy of the microstate probability distribution whose density function is $f(q,p,t)$, or in short, {\it information entropy}. The uncertainty of the first event is given by the path information entropy $S_{I}^{F} (t_0)$ defined in (\ref{eq29}). For $S_{I}^{G}(t,t_0)$, $S_{I}^{f} (t)$ and $S_{I}^{F} (t_0)$ the aforementioned property of information entropies is given here by the following relation: 
\begin{equation}
S_{I}^{G}(t,t_0) \leq  S_{I}^{f} (t) + S_{I}^{F} (t_0) , \label{eq32}
\end{equation}
with the equality if (and only if) the two random variables defining the individual events are independent. Furthermore, from (\ref{eq30}) and (\ref{eq32}), one obtains an important relation between the information entropy $S_{I}^{f} (t)$ and the conditional information entropy $S_{I}^{DF}(t,t_0)$:
\begin{equation}
S_{I}^{f} (t) \geq  S_{I}^{DF}(t,t_0) , \label{eq33}
\end{equation}
with the equality if (and only if) the two random variables defining the individual events are independent. In the case when Hamilton's equations depend on time, our analysis is based on the joint density $\mathcal{F}(q,p,t; q_0,p_0 ,t_0)$. Following the same argumentation leading to (\ref{eq33}), one obtains the relation between the information entropy $S_{I}^{f} (t)$ and the conditional information entropy $S_{I}^{Bf}(t,t_0)$: 
\begin{equation}
S_{I}^{f} (t) \geq  S_{I}^{Bf}(t,t_0) , \label{eq33a}
\end{equation}
with the equality only in the case of independence of the two random variables.

In terms of probability, the events occurring at time $t_0$ on the set $\Omega (M)$ of all possible phase space paths and at any time $t$ on the set $M \subset \Gamma $ of all possible phase space points, are not independent. If we assume that at given initial time $t = t_0$ the values of joint probability density $G(q,p,t;(q_0,p_0)_\omega ,t_0) $ are physically well defined (in the sense of (\ref{eq18})) for all points $(q,p) \in \Gamma $ and $(q_0,p_0) \in S_0(M)$, then its values are determined at all times $t$ in the entire phase space $\Gamma$ via the Liouville equation (\ref{eq24}). Simple deduction leads to the conclusion that maximization of the conditional information entropy $S_{I}^{DF} (t, t_0)$, subject to the constraints of Liouville equation (\ref{eq24}) and normalization, can not attain the upper bound which is given (at any time $t$) by the value of the information entropy $S_{I}^{f} (t)$ in (\ref{eq33}). Attaining this upper bound would require statistical independence, which would have as its logical consequence a complete loss of correlation between the paths in the set $\Omega (M)$ of possible phase space paths at time $t_0$ and the points in the set $M \subset \Gamma $ of possible phase space points at time $t$. Statistical independence is precluded at any time $t$ by the constraint implied by the Liouville equation (\ref{eq24}), and the requirement that the joint probability density $G(q,p,t;(q_0,p_0)_\omega ,t_0) $ is well defined. 

If  the conditional information entropy $S_{I}^{Bf} (t, t_0)$ is maximized subject to the constraints of Liouville equation (\ref{eq24c}) for the joint probability density $\mathcal{F}(q,p,t; q_0,p_0 ,t_0)$ and normalization, by similar deduction the same conclusion is obtained for $S_{I}^{Bf} (t, t_0)$ and its upper bound given by (\ref{eq33a}). Furthermore, statistical independence between phase space points at time $t_0$ and $t$ implies statistical independence between phase space paths at time $t_0$ and phase space points at time $t$. The converse, on the other hand, is not always true. A phase space path consists of infinitely many points. In the case of time independent Hamiltonian function phase space path is specified uniquely by all these points independently of time. In that case, therefore, statistical independence between phase space paths at time $t_0$ and phase space points at time $t$ is not sufficient for the statistical independence between points at time $t_0$ and $t$.

\section{A model for a closed Hamiltonian system}\label{secAMFCHS}

At this point it is helpful to make a distinction between two different aspects of time evolution. The first is a microscopic aspect which represents a problem of dynamics implied in this work by Hamilton's equations. The solutions are represented in $\Gamma $ as phase space paths. Predicting macroscopic time evolution represents a problem of available information and inferences from that partial information. Therefore, along with the microscopic state which is never known completely, microscopic dynamics and the respective phase space paths are also part of this problem of incomplete information. In the case of macroscopic system information about microscopic dynamics is very likely to be incomplete for variety of different possible reasons. Some of them will be analyzed in the subsequent paper \cite{kuic2}. However, in the absence of more complete knowledge, Hamilton's equations (\ref{eq1}) and the set of possible phase space paths are the representation of our prior information about microscopic dynamics. It is natural to assume that the macroscopic time evolution which we are trying to predict is consistent with our knowledge of microscopic dynamics, even when this knowledge is not complete. 

All arguments mentioned before lead to the conclusion that regarding Liouville equation (\ref{eq24}) as a strict {\it microscopic constraint} on time evolution in terms of prediction is equivalent to having complete information about microscopic dynamics. Following the previously introduced assumptions, the Liouville equation (\ref{eq24}) can also be regarded as a {\it macroscopic constraint} on time evolution. If our information about microscopic dynamics is not sufficiently detailed to completely determine the time evolution, an average is taken over all cases which are possible on the basis of partial information. In predictive statistical mechanics formulated by Jaynes, inferences are drawn from probability distributions whose sample spaces represent what is known about the structure of microstates, and maximize information entropy subject to the available macroscopic data as constraints \cite{jaynes6}. In this way ``objectivity" of probability assignments and predictions is ensured from introducing additional assumptions which are not necessarily  contained in the available data. In the simple model developed in \cite{kuic} we have introduced the same basic idea into stage (2) (explained in Section \ref{eqI}) of the problem of prediction for closed Hamiltonian systems. The conditional information entropy $S_{I}^{DF} (t, t_0)$ is maximized subject to the constraint of Liouville equation (\ref{eq24}), introduced as a phase space average, or more precisely, an integral over phase space similarly to other macroscopic constraints.  This approach allowed us to consider the incomplete nature of our information about microscopic dynamics in a rational way, and leads to the loss of correlation between the initial phase space paths and final microstates and to corresponding uncertainty in prediction. The conditional information entropy $S_{I}^{DF} (t, t_0)$ is the measure of this uncertainty, related to loss of information about the state of the system.

Now we present very briefly the basic theoretical model of the macroscopic time evolution of closed Hamiltonian systems which is the basis for further generalizations that will be introduced in the subsequent paper \cite{kuic2}. Details of this model were partially presented in \cite{kuic}. In the first approach to this basic model, time evolution of the conditional probability density $D(q,p,t | (q_0,p_0)_\omega ,t_0 )$ in the interval $t_0 \leq t \leq t_a$ is determined from the maximization of the conditional information entropy $S_{I}^{DF} (t, t_0)$ under the following two constraints: normalization condition  
\begin{equation}
\int _M D(q,p,t | (q_0,p_0)_\omega ,t_0 )d\Gamma  = 1 , \label{eq37}
\end{equation}
and the Liouville equation for $D(q,p,t | (q_0,p_0)_\omega ,t_0 )$,
\begin{equation}
{\partial D \over \partial t} + \sum _{i=1} ^s \left ({\partial D \over \partial q_i}{\partial H \over \partial p_i} - {\partial D \over \partial p_i}{\partial H \over \partial q_i}\right ) = 0 . \label{eq36}
\end{equation} 
From (\ref{eq23}) it follows that the constraints given by (\ref{eq24}) and (\ref{eq36}) are equivalent. By definition, the set of all possible microstates $M \subset \Gamma$ is an invariant set. The normalization constraint (\ref{eq37}) contains information about the structure of possible microstates in $\Gamma $, in the time interval under consideration $t_0 \leq t \leq t_a$. Information about microscopic dynamics is represented by Hamilton's equations (\ref{eq1}) and the set $\Omega (M)$ of possible phase space paths in $\Gamma$. In addition, this information is also contained in the Liouville equation (\ref{eq36}). By introducing the Liouville equation for $D(q,p,t | (q_0,p_0)_\omega ,t_0 )$ as a strict microscopic constraint (\ref{eq36}) the time evolution  is completely determined by this equation and initial conditions. Maximization of the conditional information entropy $S_{I}^{DF} (t, t_0)$ subject to this constraint and the normalization is therefore equivalent to solving the Liouville equation for $D(q,p,t | (q_0,p_0)_\omega ,t_0 )$ that maximizes it. This approach was introduced in \cite{kuic} to prove the consistency of the basic model and therefore will not be exposed further in the current paper. As was already explained in Sect. \ref{secIE}, for any physically well defined conditional probability density $D(q,p,t | (q_0,p_0)_\omega ,t_0 )$ (in the sense of (\ref{eq21}) and (\ref{eq23})), the upper bound on $S_{I}^{DF}(t,t_0)$, given by (\ref{eq33}), is not attained in the maximization under constraints (\ref{eq37}) and (\ref{eq36}). In this approach to the basic theoretical model there is no statistical independence between the initial phase space paths and final microstates. Furthermore, the value of $S_{I}^{DF}(t,t_0)$ is constant during the time interval under consideration $t_0 \leq t \leq t_a$ and there is no loss of information about the state of the system. 
 
The conclusions which follow from the interpretation of relation (\ref{eq33}) and the property of $S_{I}^{DF}(t,t_0)$ as a measure of uncertainty related to loss of information were argumented in Sect. \ref{secIE}. In the second approach (given also in \cite{kuic}) to this basic model, these conclusions are taken into account by replacing the strict equality constraint (\ref{eq36}) with the constraint in the form of the integral over phase space,
\begin{equation}
\varphi _2 ((q_0,p_0)_\omega ,t_0; t, D) =  \int_M \left [{\partial D \over \partial t} + \sum _{i=1} ^s \left ({\partial D \over \partial q_i}{\partial H \over \partial p_i} - {\partial D \over \partial p_i}{\partial H \over \partial q_i}\right )\right ]F \ d\Gamma = 0 . \label{eq63}
\end{equation}
The normalization constraint (\ref{eq37}) is writen here in equivalent but more suitable form:
\begin{eqnarray}
\varphi_1 ((q_0,p_0)_\omega ,t_0; t, D) = F \int_M D \ d\Gamma   -  F = 0 . \label{eq43}
\end{eqnarray}  
Time derivative of the conditional information entropy $S_{I}^{DF}(t,t_0)$ given by (\ref{eq28}) is equal to
\begin{equation}
\frac {dS_{I}^{DF}(t,t_0)}{dt} = - \int _{S_0(M)} \int_M \frac{\partial D}{\partial t} F\log D \ d\Gamma dS _0 -  \int _{S_0(M)} \int_M \frac{\partial D}{\partial t} F \ d\Gamma dS _0 . \label{eq38}
\end{equation}
Because of the normalization (\ref{eq37}), the last term in (\ref{eq38}) is equal to zero. At time $t_a$ conditional information entropy $S_{I}^{DF}(t_a,t_0)$ is given by the expression, 
\begin{equation}
S_{I}^{DF} (t_a, t_0) = - \int _{t_0}^{t_a} \int _{S_0(M)} \int_M \frac{\partial D}{\partial t}F\log D \ d\Gamma dS _0 dt + S_{I}^{DF}(t_0, t_0) . \label{eq40}
\end{equation}
It is suitable to form the following functional 
\begin{equation}
J[D] = S_{I}^{DF} (t_a, t_0) - S_{I}^{DF}(t_0, t_0) =  \int _{t_0}^{t_a} \int _{S_0(M)} \int_M  K(D, \partial _t D) d\Gamma dS _0 dt , \label{eq41}
\end{equation}
with the function $K(D, \partial _t D)$ given by
\begin{equation}
K(D, \partial _t D) = - \frac{\partial D}{\partial t}F \log D . \label{eq42}
\end{equation} 
In the variational problem which is considered here, functional $J[D]$ in (\ref{eq41}) is rendered stationary with respect to variations subject to the constraints (\ref{eq63}) and (\ref{eq43}). The prescribed $D(q,p,t | (q_0,p_0)_\omega ,t_0 )$ at initial time $t_0$ must be physically well defined in the sense of (\ref{eq21}) and (\ref{eq23}).  In this variational problem, function $D(q,p,t | (q_0,p_0)_\omega ,t_0 )$  is not required to take on prescribed values on the remaining portion of the boundary of integration region $M \times (t_0,t_a)$ in (\ref{eq40}).

Methods for variational problems with this type of constraints exist and one can develop them and apply in practical problems \cite{wan}. Here, in the notation which is adapted to this particular problem, the following functionals are introduced:      
\begin{equation}
C_1[D, \lambda _1] = \int_{S_0(M)} \int _{t_0}^{t_a} \lambda _1 \varphi_1 \ dt dS_0  , \label{eq45}
\end{equation} 
and
\begin{equation}
C_2[D, \lambda _2] = \int_{S_0(M)} \int _{t_0}^{t_a} \lambda _2\varphi _2 \ dt dS_0. \label{eq64}
\end{equation} 
The Lagrange multipliers $\lambda _1 \equiv \lambda _1 ((q_0,p_0)_\omega ,t_0; t)$ and $\lambda _2 \equiv \lambda _2((q_0,p_0)_\omega ,t_0; t)$ are functions defined in the integration regions in (\ref{eq45}) and (\ref{eq64}). For any function with continuous first partial derivatives, Euler equation for the constraint $\varphi _2 \equiv \varphi _2 ((q_0,p_0)_\omega ,t_0; t, D)$ is equal to zero. Following the most general multiplier rule for this type of problems which is explained in detail in ref. \cite{wan}, we introduce an additional constant Lagrange multiplier $\lambda _0$ for the function $K$, 
\begin{equation} 
J[D, \lambda_0] = \int _{t_0}^{t_a} \int _{S_0(M)} \int_M  \lambda _0 K(D, \partial _t D) \ d\Gamma dS _0 dt . \label{eq47}
\end{equation}
The functional $I[D, \lambda_0, \lambda_1, \lambda_2 ]$ is formed from $J[D, \lambda_0]$, $C_1[D, \lambda _1]$ and $C_2[D, \lambda _2]$:  
\begin{equation}
I[D, \lambda_0, \lambda_1, \lambda_2 ] = J[D, \lambda_0] - C_1[D, \lambda _1] - C_2[D, \lambda _2] \label{eq48} . 
\end{equation} 
The existence of Lagrange multipliers $\lambda_0 \ne 0$, and $\lambda_1$, $\lambda_2 $ not all equal to zero, such that the variation of $I[D, \lambda_0, \lambda_1, \lambda_2 ]$ is stationary  $\delta I = 0$, represents a proof that it is possible to make $J[D]$ in (\ref{eq41}) stationary subject to constraints (\ref{eq63}) and (\ref{eq43}). For a function $D(q,p,t | (q_0,p_0)_\omega ,t_0 )$ to maximize $S_{I}^{DF}(t_a,t_0)$ subject to constraints (\ref{eq63}) and (\ref{eq43}), it is necessary that it satisfies the Euler equation: 
\begin{eqnarray}
\lambda _0 \left \{\frac{\partial K}{\partial D} - \frac{d}{dt}\left (\frac{\partial K}{\partial (\partial _t D)}\right ) - \sum_ {i = 1}^s \left [\frac{d}{dq_i}\left (\frac{\partial K}{\partial (\partial _{q_i} D)}\right ) + \frac{d}{dp_i}\left (\frac{\partial K}{\partial (\partial _{p_i} D)}\right ) \right ]\right \} - \lambda _1F + {\partial \lambda _2 \over \partial t} F   = 0 . \label{eq65}
\end{eqnarray}
It is easy to check that the term multiplied by $\lambda _0$ in Euler equation (\ref{eq65}) is equal to zero. Stationarity of the functional $I[D, \lambda_0, \lambda_1, \lambda_2 ]$ in (\ref{eq48}) is therefore possible even with $\lambda _0 \ne 0$. From (\ref{eq65}) it follows that the Lagrange multipliers $\lambda_1$ and $\lambda_2$ satisfy the equation
\begin{equation}
{\partial \lambda _2 \over \partial t} =  \lambda _1  . \label{eq66}
\end{equation}

Another necessary condition for a maximum, in addition to (\ref{eq65}), exists if function $D(q,p,t | (q_0,p_0)_\omega ,t_0 )$ is not required to take on prescribed values on a portion of the boundary of $M \times (t_0,t_a)$: then, it is necessary that $D(q,p,t | (q_0,p_0)_\omega ,t_0 )$ also satisfies the Euler boundary condition on the portion of the boundary of $M \times (t_0,t_a)$ where its values are not prescribed, ref. \cite{wan}. In accordance with this, for all points on the portion of the boundary of $M \times (t_0,t_a)$ where $t = t_a$, the Euler boundary condition gives: 
\begin{equation}
\left [\frac{\partial K}{\partial (\partial _t D)} - \lambda _2F\right ] _{t = t_a}  = - \left [ \log D + \lambda _2\right ]_{t = t_a} F = 0 . \label{eq68}
\end{equation}
For all points on the portion of the boundary of $M \times (t_0,t_a)$ where time $t$ is in the interval $t_0 < t < t_a$, the Euler boundary condition gives:
\begin{equation}
F \left [\lambda _2 {\bf v} \cdot {\bf n}\right ]_{\ \mathrm {at \ the \ boundary \ of} \ M}  = 0 . \label{eq55}
\end{equation}
In (\ref{eq55}), ${\bf v} \cdot {\bf n}$ is a scalar product of the velocity field ${\bf v} (q,p)$ in $\Gamma $ (defined in Sect. \ref{secHD}) and the unit normal ${\bf n}$ of the boundary surface of invariant set $M$, taken at the surface. Equation (\ref{eq55}) is satisfied naturally due to Hamiltonian motion, since the set $M$ is invariant by definition, and therefore  ${\bf v} \cdot {\bf n} = 0$ for all points on the boundary surface of $M$. This is a consequence of the fact that phase space paths do not cross over the boundary surface of the invariant set $M$.

The form of the MaxEnt conditional probability density at time $t_a$ follows from (\ref{eq68}):  
\begin{equation}
D(q,p,t_a | (q_0,p_0)_\omega ,t_0 )  = \exp \left [ - \lambda _2 ((q_0,p_0)_\omega ,t_0; t_a) \right ] . \label{eq70}
\end{equation}
For any, at initial time $t_0$ well defined conditional probability density, there is an entire class of equally probable solutions $\{D(q,p,t | (q_0,p_0)_\omega ,t_0 )\}$ obtained by MaxEnt algorithm, which all satisfy the macroscopic constraint (\ref{eq63}). At time $t_a$, all functions in this class of MaxEnt solutions are equal and given by (\ref{eq70}). With the exception of times $t_0$ and $t_a$, the conditional probability density $D(q,p,t | (q_0,p_0)_\omega ,t_0 )$ obtained by MaxEnt algorithm is not uniquely determined in the interval $t_0 < t < t_a$. This is a consequence of the fact that the macroscopic constraint (\ref{eq63}) does not determine the time evolution of $D(q,p,t | (q_0,p_0)_\omega ,t_0 )$ uniquely, in the way that the strict microscopic constraint (\ref{eq36}) does. However, MaxEnt solutions still predict only time evolutions entirely within the invariant set $M$, due to (\ref{eq55}). This property follows from the constraint (\ref{eq63}), and takes into account the information about the constants of motion that determine the invariant set $M$, and in that way, about related conservation laws.

From the normalization (\ref{eq37}) of the conditional probability density, given at time $t_a$ by (\ref{eq70}), one obtains the relation:  
\begin{equation}
W(M) \exp \left [ - \lambda _2 ((q_0,p_0)_\omega ,t_0; t_a) \right ] = 1 , \label{eq70a} 
\end{equation}
where $W(M)$ is the measure, i.e., phase space volume of the invariant set $M$. Equation (\ref{eq70a}) implies that the Lagrange multiplier $\lambda _2 ((q_0,p_0)_\omega ,t_0; t)$ at time $t = t_a$ is independent of the variables $(q_0,p_0)_\omega $:
\begin{equation}
\lambda _2 ((q_0,p_0)_\omega ,t_0; t_a) = \lambda _2 (t_a) . \label{eq71}
\end{equation}
Microstate probability density $f(q,p,t)$ at time  $t = t_a$ is then calculated by using: (\ref{eq17}) and (\ref{eq23}), the MaxEnt conditional probability density $D(q,p,t | (q_0,p_0)_\omega ,t_0 )$ at time  $t = t_a$ given by (\ref{eq70}) and (\ref{eq71}), and the path probability distribution $F((q_0,p_0)_\omega ,t_0 )$ at initial time $t_0$:
\begin{equation}
f(q,p,t_a) = \exp \left [ - \lambda _2 (t_a) \right ] . \label{eq72}
\end{equation}
It follows from (\ref{eq70}--\ref{eq72}) that at time  $t_a$, the MaxEnt conditional probability density and the corresponding microstate probability density are equal,
\begin{equation}
D(q,p,t_a | (q_0,p_0)_\omega ,t_0 ) = f(q,p,t_a) = \exp \left [ - \lambda _2 (t_a) \right ] = \frac{1}{W(M)} . \label{eq73}
\end{equation}
From (\ref{eq28}), (\ref{eq31}) and (\ref{eq73}), one obtains the values of information entropies $S_{I}^{DF}(t,t_0)$ and $S_I ^f (t)$  at time $t_a$,
\begin{equation}
S_I ^f (t_a) = S_{I}^{DF}(t_a, t_0) = \log W(M) . \label{eq74} 
\end{equation}

Equalities (\ref{eq73}) and (\ref{eq74}) are possible only in case of statistical independence. Logical consequence of the statistical independence is the complete loss of correlation between the phase space paths at time $t_0$, and the microstates at time $t_a$. In general, property of macroscopic systems is that they appear to randomize themselves between observations, provided that the observations follow each other by a time interval longer then a certain characteristic time $\tau $ called the relaxation time \cite{kittel}. In the interpretation given here, relaxation time $\tau $ for a closed Hamiltonian system represents a characteristic time required for the described loss of correlation between the initial phase space paths and final microstates. Furthermore, $\tau $ also represents a time interval during which predictions, based on incomplete information about microscopic dynamics, become uncertain to a maximum extent compatible with the available data. This uncertainty is related to loss of information about the state of the system. 

This interpretation is reflected in the role of the Lagrange multipliers $\lambda _1 ((q_0,p_0)_\omega ,t_0; t)$ and $\lambda _2 ((q_0,p_0)_\omega ,t_0; t)$. They are required to satisfy (\ref{eq66}), and by integrating it one obtains the following relation, 
\begin{equation}
 \lambda _2 ((q_0,p_0)_\omega ,t_0; t) = \int_{t_0}^{t} \lambda _1 ((q_0,p_0)_\omega ,t_0; t^{\prime})dt^{\prime} + \lambda _2 ((q_0,p_0)_\omega ,t_0; t_0) , \label{eq75}
\end{equation} 
for all $t$ in the interval $t_0 \leq  t \leq  t_a$. By using (\ref{eq75}), with (\ref{eq70a}), (\ref{eq71}) and (\ref{eq74}), one obtains
\begin{equation}
 S_I ^f (t_a) = S_{I}^{DF}(t_a,t_0) = \log W(M) = \int_{t_0}^{t_a} \lambda _1 ((q_0,p_0)_\omega ,t_0; t)dt + \lambda _2 ((q_0,p_0)_\omega ,t_0; t_0) . \label{eq77}
\end{equation} 
It is clear, from relations (\ref{eq71}), (\ref{eq75}) and (\ref{eq77}), that at time $t_a$ the Lagrange multiplier $\lambda _2 ((q_0,p_0)_\omega ,t_0; t_a ) \equiv \lambda _2 (t_a)$ is determined by the measure $W(M)$ of the invariant set $M$ of all possible microstates, i.e., the volume of accessible phase space. The subsequent application of MaxEnt algorithm of the described type for a closed system with Hamiltonian dynamics, without the introduction of additional constraints, results in the increase of  $W(M)$. From (\ref{eq71}), (\ref{eq75}) and (\ref{eq77}) it is then deduced that $\lambda _2(t_a) \geq  \lambda _2(t_0)$.

Information about the structure of possible microstates restricts the corresponding set, and therefore sets an upper bound on the volume of accessible phase space. The values of $S_{I}^{DF}(t_a,t_0)$ and $S_I ^f (t_a)$ at time $t_a$, given in (\ref{eq77}), are equal to the maximum value of the {\it Boltzmann-Gibbs entropy}, compatible with this information. The Lagrange multiplier $\lambda _1 ((q_0,p_0)_\omega ,t_0; t)$, integrated in (\ref{eq77}) over time $t_0 \leq t \leq t_a$, is then determined by the rate at which the maximum Boltzmann-Gibbs entropy is attained in a reproducible time evolution. The integral in (\ref{eq77}), and the quantity $\lambda _1 ((q_0,p_0)_\omega ,t_0; t)$, can be identified with the change in entropy, and the {\it rate of entropy change} for a closed Hamiltonian system, respectively. If the information about microscopic dynamics of a closed Hamiltonian system is considered complete, whether entropy production can be defined without recourse to coarse graining procedures, or macroscopic, phenomenological approaches, remains an open question. In general, part of information is discarded in all such models, at some stage, in order to match with what is observed in nature in various manifestations of the second law of thermodynamics. The model developed in \cite{kuic} corresponds to a closed system with the time independent Hamiltonian function. The model presented in the current paper includes also closed systems with Hamiltonian function that depends on time, and the same conclusions are obtained analogously. In that case, model is modified by a simple change of the symbols with corresponding meanings, as explained in Sections \ref{sec_Mic} and \ref{secIE}: replace $F((q_0,p_0)_\omega ,t_0 )$ with $f(q_0,p_0, t_0)$, $D(q,p , t \vert (q_0,p_0)_\omega , t_0)$ with $B(q,p , t \vert q_0,p_0 , t_0)$, $M$ and $S_0(M)$ with $\Gamma $, $dS_0$ with $d\Gamma _0$, and $S_{I}^{DF} (t, t_0)$ with $S_{I}^{Bf} (t, t_0)$.

\section{Conclusion}\label{secCGARWNT}

It is demonstrated that Jaynes' interpretation of irreversibility as a consequence of a gradual loss of information as to the state of the system due to our inability to follow its exact time evolution during the process \cite{jaynes2}, has a clear mathematical formulation in the concepts which are introduced in this paper. The most important theoretical concept in this work was the maximization of the conditional information entropy subject to given constraints, and its relation with the information entropy, taken from Shannon's information theory. At a same time, the key element of this theoretical approach was the introduction of Liouville equation for the conditional probability distribution as a macroscopic constraint, i.e., as a  constraint given by averaging this equation in the integral over the available phase space. In this way, in the problem of predicting the macroscopic time evolution of closed Hamiltonian systems, the incompleteness of our information about the detailed microscopic dynamics of the system is included, in a way which is consistent with the foundational principles of predictive statistical mechanics. It is demonstrated that such mathematical description results in a total loss of correlation between the initial phase space paths and final microstates. This loss of correlation is related to a loss of information about possible microstates of the system, which is brought into connection with the change of entropy of the system. This connection allowed the definition of the entropy change and the rate of entropy change for a closed Hamiltonian system without additional assumptions. In the subsequent paper \cite{kuic2}, we show, by generalizing this approach and including, as the additional constraints, the relevant information for prediction of macroscopic time evolution on the hydrodynamic time scale, that it is consistent with the known results of the nonequilibrium statistical mechanics and thermodynamics of irreversible processes.

\end{document}